\documentclass{article}

\usepackage{arxiv}

\usepackage[utf8]{inputenc} 
\usepackage[T1]{fontenc}    
\usepackage{hyperref}       
\usepackage{url}            
\usepackage{booktabs}       
\usepackage{amsfonts}       
\usepackage{nicefrac}       
\usepackage{microtype}      

\usepackage{amsmath,amsfonts,bm}
\usepackage{graphicx}
\usepackage{textcomp}
\usepackage{xcolor}
\usepackage{enumitem}
\usepackage{graphicx}
\usepackage{multirow}
\usepackage[linesnumbered,ruled,noend]{algorithm2e}
\usepackage{parskip}
\usepackage{xspace}

\SetAlgoCaptionSeparator{:}
\SetNlSty{texttt}{}{:}
\usepackage[T1]{fontenc}
\usepackage[font=small,labelfont=bf,tableposition=top]{caption}

\DeclareCaptionLabelFormat{andtable}{#1~#2  \&  \tablename~\thetable}

\SetAlgoCaptionSeparator{:}
\SetNlSty{texttt}{}{:}

\newcommand{\eg}{\emph{e.g.}}
\newcommand{\ie}{\emph{i.e.}}
\newcommand{\first}[1]{\mathit{first}({#1})}
\newcommand{\last}[1]{\mathit{last}({#1})}
\newcommand{\wrt}{ with respect to }
\newcommand{\flowminer}{\emph{FlowMiner}\xspace}
\newcommand{\perracotta}{\emph{Perracotta}\xspace}

\newtheorem{definition}{Definition}[section]

\title{Mining Message Flows from System-on-Chip Execution Traces}

\author{
  MD Rubel Ahmed,~Hao Zheng \\
  Department of Computer Science and Engineering\\
  University of South Florida\\
  Tampa, FL 33620 \\
  \texttt{\{mdrubelahmed, haozheng\}@usf.edu} \\
   \And
 Parijat Mukherjee \\
Intel \\
  Hillsboro, Oregon \\
  \texttt{Parijat.Mukherjee@intel.com} \\
\And
Mahesh C. Ketkar \\
Intel \\
  Folsom, California \\
  \texttt{Parijat.Mukherjee@intel.com} \\
  \And
  Jin Yang \\
Intel \\
  Hillsboro, Oregon \\
  \texttt{Parijat.Mukherjee@intel.com} \\
}

\begin{document}
\maketitle

\begin{abstract}
Comprehensive and well-defined specifications are necessary to perform rigorous and thorough validation of system-on-chip (SoC) designs.  Message flows specify how components of an SoC design communicate and coordinate with each other to realize various system functions. Message flow specifications are essential for efficient system-level validation and debug for SoC designs. However, in practice such specifications are usually not available, often ambiguous, incomplete, or even contain errors. This paper addresses that problem by proposing a specification mining framework, \emph{FlowMiner}, that automatically extracts message flows from SoC execution traces, which, unlike software traces, show a high degree of concurrency. A set of inference rules and optimization techniques are presented to improve mining performance and reduce mining complexity. Evaluation of this framework in several experiments shows promising results.
\end{abstract}


\section{Introduction}
\label{motivation}
System-on-chip (SoC) designs integrate a large number of functional blocks that communicate and coordinate with each other to implement various sophisticated functions according to system-level protocols. Experiences have shown that the implementation of those system-level protocols is the major source of various design errors. Therefore, communication-centric validation and debug methods have attracted a lot of attentions recently.

Specifications on system-level communications are essential for communication-centric validation and debug methods. In practice, such specifications are usually not available, often ambiguous, incomplete, or even contain errors. Very often, such specifications can also become outdated and disconnected from the design implementation as the design progresses.

\begin{figure}[tb]

\begin{center}
\includegraphics[width=.5\textwidth,angle=0]{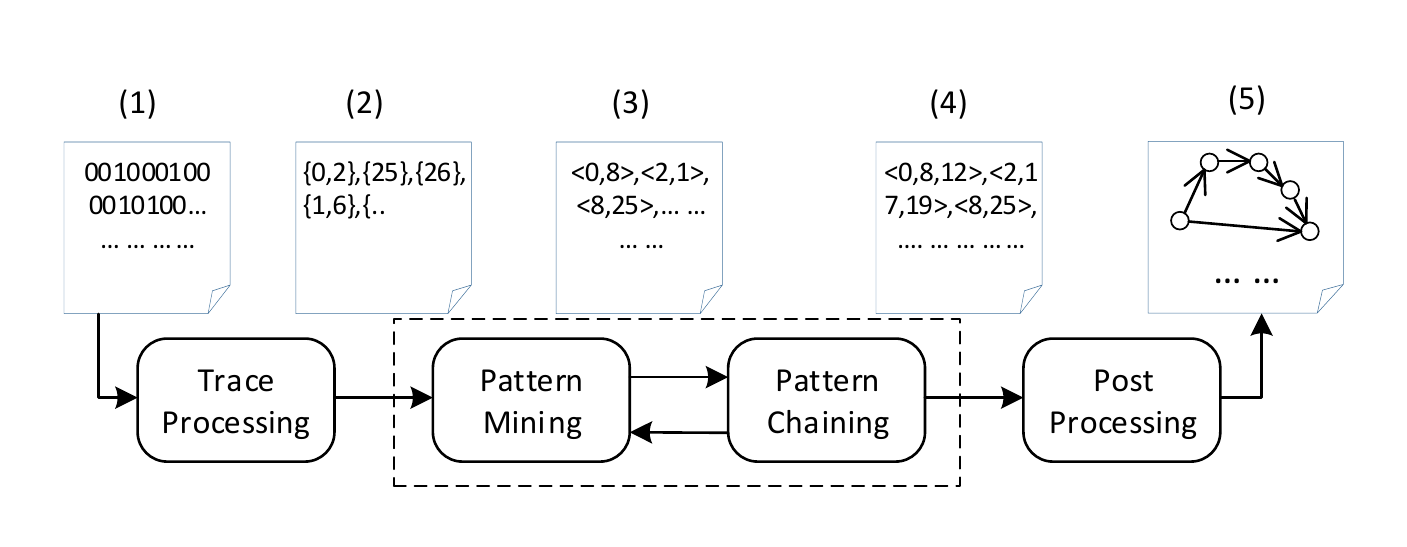}
\caption{The overview of \textit{FlowMiner} for specification mining from SoC execution traces. (1) Collect SoC execution traces (2) Process traces for mining, (3--4) $FlowMiner$ to extract sequential patterns from SoC execution traces, (5) Post-processing of mined patterns.}
\label{workflow}
\end{center}
\vspace{-7mm}
\end{figure}

Due to the nature of SoC designs, the SoC executions are highly concurrent where multiple system transactions are executed simultaneously.  Therefore, the existing software trace mining methods are inadequate to handle SoC execution traces.
In this paper, we present a specification mining framework, \textit{FlowMiner}, to address the above problem.  The overview of \textit{FlowMiner} is shown in Fig.~\ref{workflow}. 

It takes as input a set of execution traces over messages observed in various communication links in a SoC design, and produces a set of a set of sequential patterns that satisfy certain interestingness measures and other requirements.  Input traces first go through a pre-processing step where a couple of trace slicing techniques are applied for better mining performance.  Next, the traces go through the actual mining process which consists of a basic mining procedure and several techniques for discovering longer patterns.  In the final step, mined patterns are processed to remove redundancy, and a visual representation of these patterns is produced for user examination.

The \textbf{contributions} of this paper are 
\begin{itemize}
\item The \emph{FlowMiner}, a new specification mining framework that can automatically infer sequential patterns about the system level communication behavior from highly concurrent SoC execution traces
\item A set of inference rules that allow longer sequential patterns to be extracted without the need of the expensive mining step, thus reducing complexity greatly, 
\item Comprehensive evaluation to demonstrate its effectiveness in comparison with existing tools. 
\end{itemize}

The organization of this paper is as follows. Section II surveys related literature in specification mining. Section III provides required background for related concepts and formalizes the problem. Section IV describes the proposed specification mining algorithm. Section V presents the experimental results, section VI tells about observed limitations and we conclude in section VII.

\section{Related Work}
\label{literature_review}
Specification has been a growing point of interest for mining assertions from various artifacts~\cite{Dwyer:1999:icse,Ammons:2002:popl,Chang:2010:aspdac,Li2010DAC,Hertz:2013:tcad,Danese:2015:vlsi-soc,Danese:2015:date,Danese:2017:dac}.
In \cite{Ammons:2002:popl}, a dynamic model based specification mining approach has been proposed that builds Finite State Automatons (FSA) from the execution logs. Learned FSAs can capture both temporal and data dependencies which could be used to verify API or ADT protocols. Another model based approach \textit{Synoptic} \cite{Beschastnikh:2011:LEI:2025113.2025151} mines invariants from log of sequential execution traces where concurrency is recorded in partial order. It generates a FSM that satisfies the temporal invariants mined from the logs. Such model based approaches target software traces and do not perform well for concurrent hardware traces\cite{Mrowca:2019:LTS:3316781.3317847}. A framework has been proposed in \cite{Lo:2007:ase} to infer temporal properties of a program as pre-chart and post-chart rules. The almost invariants mined by this tool represents total order of messages, so it may miss many partial order or concurrent scenarios which is common in SoC executions.
The approaches presented in~\cite{Li2010DAC,Hertz:2013:tcad,Danese:2015:vlsi-soc,Danese:2015:date,Danese:2017:dac} mine assertions from either gate-level representations~\cite{Li2010DAC}, or RTL models~\cite{Chang:2010:aspdac,Hertz:2013:tcad,Danese:2015:vlsi-soc,Danese:2015:date,Danese:2017:dac}.
Work \cite{Liu:2013} describes an assertion mining approach in the form of  sequential patterns from the simulation traces of system level designs using data mining and supervised learning algorithms. It mines assertions relying on the system level assertions mined by another tool called \textit{Goldmine}~\cite{5457129} that requires the system itself to flush out spurious assertions. This flushing may not be feasible in many cases. Aside from this, these miners only focus on assertion generation, not complex execution patterns deduction, thus not much interesting for building flow specification scenarios.
Some recent approaches are trying to solve this specification mining problem taking advantage of deep learning algorithms. As the execution traces have the temporal component involved, a recurrent neural network (RNN) based approach is presented in \cite{Le:2018:DSM:3213846.3213876} that utilizes Long Short-Term Memory (LSTM) models. But LSTM models handles one event at time and serialises them, when concurrency involved, serialization between two more events may not be possible, so this approach is limited to sequentially executed program traces. 

The work \textit{BaySpec} \cite{Mrowca:2019:LTS:3316781.3317847}
uses a dynamic mining approach to extracts formal specifications in the form of LTL formulas from trained Baysian networks. It does not relay on some user defined template, thus may produce many useful unknown patterns.

But silicon trace specification extraction can not be done using this method, because BaySpec requires functional segmentation of traces that is starting and terminating events should be known before hand, which is very hard to know for many cases thus compromises the generalization of this tool. \emph{Perracotta}~\cite{Yang06perracotta:mining} is another execution log analysis tool that mines temporal properties of in the form of API rules. Though it does not intend to infer model of the system, it shows a chaining technique that can be used to find sequential patterns. Such patterns can be useful for inferring message flows for SoCs. But this work is also limited by invariant mining, so it fails to capture flows with complex execution scenarios.

In spite of the fact that these mining methods can bring out impressive results, they are not suitable for highly complex SoC designs, which generally have multiple tasks executing in parallel in an interleaved manner, thus produce a complex set of execution traces. We propose a generalized case of specification mining where an execution trace is composed of sets of events. Events in the same set represent concurrency, thus no ordering relation is inferred among them. \textit{FlowMiner} searches for strong temporal relation among the events of different sets and constructs flow specifications in the form of sequential patterns.



\section{Background}
\subsection{Message Flow Specification}
\label{flow-spec}

\begin{figure}[tb]
\begin{center}
\includegraphics[height=3.0in,width=2.5in,angle=0]{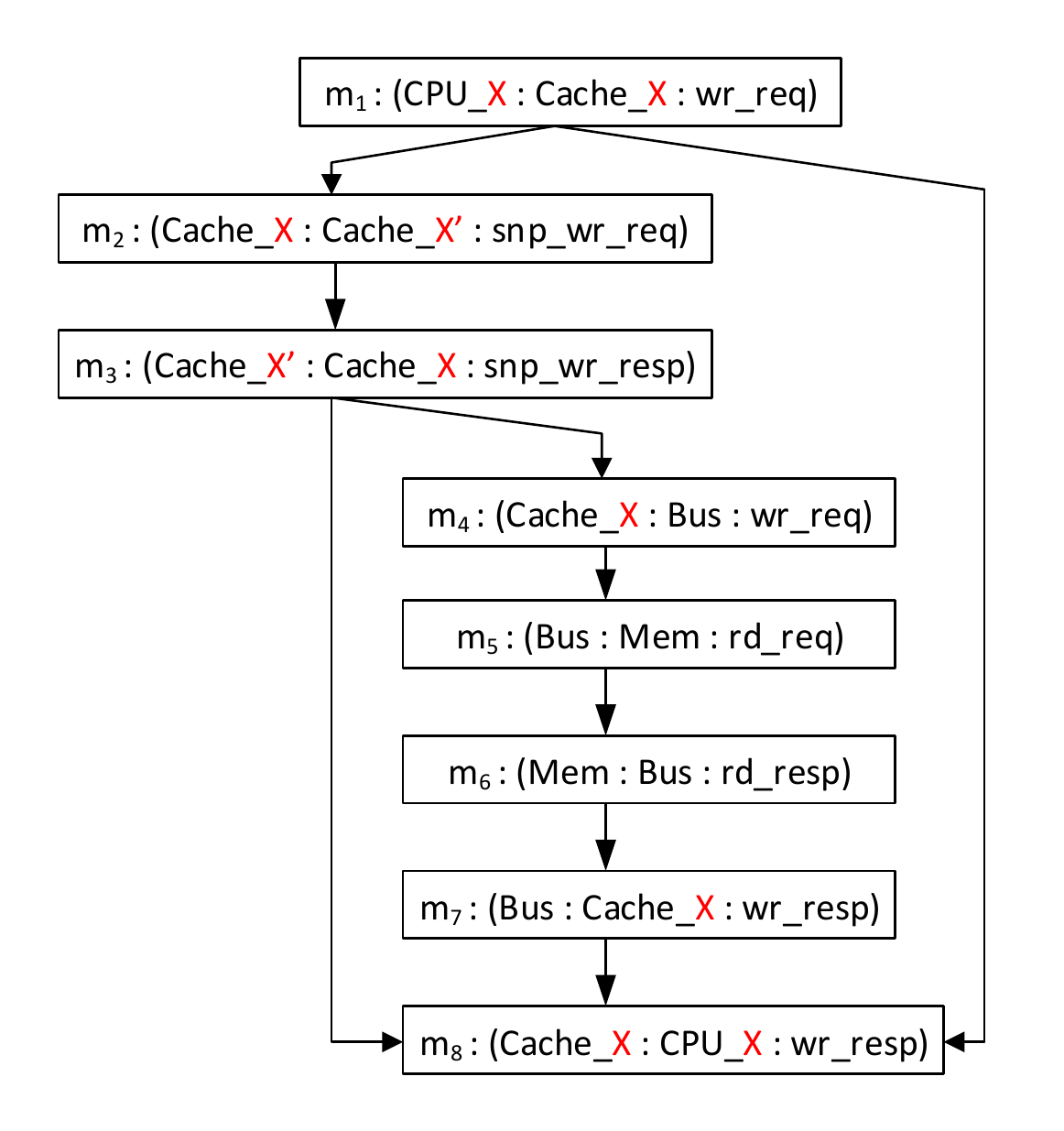}
\vspace{-10pt}
\caption{A CPU downstream write flow.}
\label{fig:ex}
\end{center}
\vspace{-7mm}
\end{figure}

System-level protocols are often represented as message flow diagrams in SoC architecture documents, therefore the word ``flow'' and ``protocol'' are used interchangeably in our context. We also refer $\textit{system flows}$ as message flows in this paper. 

A flow specification for an SoC design \textbf{system} is a collection of message flows. A message flow is represented as a directed acyclic graph with an example shown in Fig.~\ref{fig:ex}.  This flow describes how a CPU downstream write operation can be fulfilled.  Formally, a flow consists a set of messages $M$ and a set of connections $C \subseteq M\times M$.  Each message is a triple $({\tt src, dest, cmd})$ where the ${\tt src}$ denotes the originating component of the message while {\tt dest} denotes the receiving component of the message.  Field {\tt cmd} denotes the operation to be performed at {\tt dest}.  In the example shown in Fig.~\ref{fig:ex}, message {\tt (CPU\_x, Cache\_x, wr\_req)} is the write request from {\tt CPU\_x} to {\tt Cache\_x}.  Each message flow is associated with an unique start message, and may terminate with one or multiple different messages. A message flow may contain multiple branches describing different paths a system can execute such flow. For example, the flow shown in Fig.~\ref{fig:ex} has three branches covering the cases where the cache snoop operation is hit or miss.
The flow specification for an SoC is a set of flows, denoted as $\vec{F} = \{F_i\}$.

During the execution of an SoC implements a set of flows $\vec{F}$, instances of individual flows may be generated, and executed. An instance of a flow is the execution of a path of that flow where the instances of messages on that path may include runtime information, \eg memory addresses, which are typical.  Therefore, a flow can be viewed as the type of template for its instances. 

\begin{definition} Given an SoC design that implements a set of flows $\vec{F}$, an \textbf{execution} of a flow instance $\{F_{i,j}~|~F_i \in \vec{F}\}$ is a sequence of mesages $\{m_0, m_1, \ldots, m_n\}$.
\end{definition}
\noindent For example, the flow in Fig.~\ref{fig:ex} 
has three possible executions: 
\{$m_1, m_{8}$\}, \{$m_1, m_2, m_3, m_8$\}, 
and \{$m_1, m_2, m_3, m_4, m_5, m_6, m_7, m_8$\}, 
where $m_x~(1 \leq x \leq 8)$ are instances of messages shown in Fig.~\ref{fig:ex}.  

To simplify the presentation, hereafter flows~(messages) and flow instances~(message instances) are used interchangeably if their meanings are clear in the context.  

During execution of an SoC design, instances of flows it implements are executed concurrently in an interleaved manner. Typically, multiple instances of different flows are executed concurrently that are captured in the trace. Consequently, an SoC execution trace is a sequence of sets of message instances.
\begin{definition}
Suppose that an SoC design implements a flow specification $\vec{F}$. An SoC execution \textbf{trace} $\rho$ is 
\begin{center}
    $\rho = (\varepsilon_0, \varepsilon_1, \ldots, \varepsilon_n)$
\end{center}
where $\varepsilon_i = \{m_{i,0},\ldots, m_{i,k}\}$ is a set of messages observed at time $i$, and $m_{i,j} $ is an message instance of an flow instance $F_{i,j}$ for every $m_{i,j} \in \varepsilon_i$.
\end{definition}

For example, a {\tt CPU\_0} read flow and a {\tt CPU\_1} write flow might be active at time $i$.  The idea of message sets helps to represent the outcomes of these concurrently executing flows. It is also important to note that the ordering of the messages in the same set of a trace is unknown.  Therefore, given two messages $m_i$ and $m_j$ and a trace $\rho$, we define $m_i < m_j$ if $m_i \in \varepsilon_i$, $m_j \in \varepsilon_j$, and $i<j$. 

%
%
\subsection{Sequential Patterns}

In this paper, we are interested in mining flows like the CPU downstream write flow shown in Fig.~\ref{fig:ex}.  Since a flow usually consists of one or multiple executions paths, \textit{FlowMiner} aims to mine sequential patterns to characterize those execution paths, thus the entire flows. 
As these flows are implemented by the on-chip communication fabrics, the mined sequential patterns are {\em invariants} across different execution traces.  

As a convenience, for a sequence $s = (m_0, m_1, \ldots, m_n)$, we use $s[i]$ to refer to the $ith$ message in $s$, \ie~$m_i$.  We also use $s[h, i]$ to denote the sub-sequence $(m_h, \ldots, m_i)$ where $h\geq 0$, $i\leq n$, and $h \leq i$. We also use $\first{s}$ and $\last{s}$ to denote $m_0$ and $m_n$, respectively. Given two sequences $s_1 = (m_0, m_1, \ldots, m_n)$, and $s_2 = (m_{n+1}, m_{n+2}, \ldots, m_{n+k})$, $s_1\#s_2$ denotes the concatenation of $s_1$ and $s_2$, resulting in a longer sequence $(m_0, m_1, \ldots, m_{n+k})$.  Finally, we use $m$ to represent the sequence $(m)$.


    

\begin{definition}
A \textbf{sequential pattern} $p$ is a sequence of messages such that
\begin{itemize}
    \item $p$ consists of at least two different messages,
    \item All messages in $p$ are unique, and 
    \item $p$ meets the threshold of certain interestingness measure.
\end{itemize}
\end{definition}

There are many different interestingness measures defined in the previous work. See~\cite{Lo:2015} for a comprehensive list.  This work uses the ones defined below.

\begin{definition}
The {\bf support} of a message $m$ \wrt to a trace $\rho$, denoted as $\mathit{supp}(m, \rho)$, is the count of the instances of $m$ found in $\rho$.
\end{definition}

\begin{definition}
\label{fconf}
Given two sequences $s_1$, and $s_2$, the \textbf{forward confidence} of $s_2$ given $s_1$ \wrt~a trace $\rho$ is the ratio between the support of $s_1\#s_2$ and the support of $s_1$. Formally,
\[
conf_f(s_1, s_2, \rho) = \frac{supp(s_1\#s_2, \rho)}{supp(s_1, \rho)}
\]
\end{definition}
Forward confidence defines the likelihood of occurrence of sequence $s_2$ after occurrence of sequence $s_1$.

\begin{definition}
\label{bconf}
Given two sequences $s_1$, and $s_2$, the \textbf{backward confidence} of $s_1$ given $s_2$ \wrt a trace $\rho$ is the ratio between the support of $s_1\#s_2$ and the support of $s_2$. Formally,
\[
conf_b(s_1, s_2, \rho) = \frac{supp(s_1\#s_2, \rho)}{supp(s_2, \rho)}
\]
\end{definition}
Backward confidence defines the likelihood of occurrence of $s_1$ before occurrence of the sequence $s_2$.  It is not used as common as the forward confidence in previous work. It is included in \emph{FlowMiner} to allow patterns with branching structures like the one shown in Fig.~\ref{fig:ex} to be mined.  For example, suppose that all three branches of the flow in Fig.~\ref{fig:ex} are executed.  In the resulting trace, the occurrence of message $m_1$ leads to either $m_2$ or $m_8$ to occur next. The forward confidence of $m_1$ given $m_2$ is not $100\%$.  However, their backward confidence is $100\%$, which allows pattern $(m_1, m_2, \ldots)$ to be mined.

In this work, we require sequential patterns have either $100\%$ forward or backward confidence as we aim to mine invariants.   

Validity of mined patterns \wrt ground truth flows GT is defined below.  To facilitate the definition, GT is a set of sequences of messages, one for each path in a flow. 
\begin{definition}
Let  $p_m$ be a mined pattern. $p_m$ is \textbf{valid} if there exists a sequence $p_t \in \mbox{GT}$ such that for every two messages $m_i$, $m_j$ in $p_m$ and $m_i < m_j$, if $m_i$, $m_j$ are also in $p_t$, then $m_i<m_j$ in $p_t$.  Otherwise, $p_m$ is \textbf{invalid}.   
\end{definition}

The above definition indicates that a valid pattern does not possess a temporal ordering between any two messages that cannot be found in all ground truth sequences. Consider the simple example below, where $p_m$ is valid with respect to a GT pattern $p_t$ as temporal dependencies between any pair of events in $p_m$ also prevails in $p_t$.
\[
\begin{array}{ll}
    p_m: & (\textcolor{red}{0, 13, 15, 23}) \\
    p_t: & (\textcolor{red}{0}, 8, 12, \textcolor{red}{13}, \textcolor{red}{15}, \textcolor{red}{23}, 24, 25)
\end{array}
\]
The flow specifications as shown in Fig.~\ref{fig:ex} also define causality relations among messages.  For example, when message $\tt(CPU\_x:Cache\_x:wr\_req)$ occurs, it causes either $\tt(Cache\_x:Cache\_x':snp\_wr\_req)$ or $\tt(Cache\_x:CPU\_x:wr\_resp)$ to happen. The goal of specification mining is to extract such causality.  We use interestingness measures as defined above to approximate the causality relations. To make such approximation more accurate, we introduce the concept of \emph{structural causality} based on the observation: \emph{any message in an SoC execution trace is an output of a component in reaction to a previous input message.}  

\begin{definition}
\label{causal}
Messages $m_i$ and $m_j$ satisfy the structural \textbf{causality} property, denoted as $\mathit{causal}(m_i, m_j)$, if 
\begin{center}
  $m_i{\tt .src} = m_j{\tt .dest}$
\end{center}
\end{definition}

The causality as defined above is referred to as \emph{structural} to differentiate from the functional causality in the flow specifications.  As an additional requirement, for a sequence $(m_0, m_1, \ldots, m_n)$ to be regarded as a pattern, every two consecutive messages in that sequence must satisfy the structural causality property. Specifically, the condition below must hold.
\[
\forall{0\leq i \leq n-1, \mathit{causal}(m_i, m_{i+1}) \mbox{ is true.}}
\]
Hereafter, we use causality to refer to the structural causality for the discussion of \emph{FlowMiner}.
\section{Mining Framework}
\label{sec:flowminer}

The mining framework shown in Fig.~\ref{workflow} is implemented in Algorithm~\ref{algo:flowminer}. 
The algorithm works as follows. It takes a set of execution traces over a set of messages as input, and produces a set of sequential patterns as output.  First, the input traces are transformed using slicing techniques. Next, the algorithm starts with mining binary patterns (patterns of two messages).  The binary patterns are then combined with a set of inference rules to generate patterns of longer lengths. In the last step, redundant patterns are removed.  A pattern is redundant if there is another pattern such that the former is either a prefix or suffix of the latter.  A visual representation is then produced to ease the analysis and understanding of the mined patterns by users.  In the following sections, the first three steps are described.

\subsection{Trace Processing}
\label{slicing}
As indicated above, SoC execution traces are highly concurrent as multiple flow instances can be in execution simultaneously at a point of time. The execution of the flow instances are interleaved, which makes correlation among messages a highly challenging task.  Therefore, inference of a large number of invalid patterns is a common problem in traditional mining approaches.  The work in~\cite{cao_mining_2020} presents two slicing techniques to break a long trace into a set of shorter sub-traces where messages are correlated more accurately.  In \emph{FlowMiner}, function $\mathit{TraceProcessing}$ implements the address-based slicing technique in~\cite{cao_mining_2020} to achieve better mining performance.  In the following, we briefly describe that slicing technique.  Readers are referred to \cite{cao_mining_2020} for more details. 

In SoC execution traces, the message instances often contain runtime information in addition to the static information as defined in the flow specification.  One such information is memory addresses, which are common in memory related operations, \eg~CPU initiated memory reads or writes.  Therefore, if a flow instance is initiated with a memory address, all message instances of that flow instance have the same address.  
In other words, if two messages carry different addresses, they should not be correlated, and thus should be separated during mining.  The example below illustrates the main idea of the address-based trace slicing.
Suppose a trace over 3 messages $\{m_0, m_1, m_2\}$ for the slicing is shown below.
\begin{center}
    $(\{m_1(10)\},\{m_2(10),m_1(15)\}, \{m_3(15)\})$
\end{center}
In the above trace, each message instance carries a memory address in parentheses.  After the slicing, the trace is broken into two sub-traces as below.
\[
\begin{array}{ll}
    addr  =  10: & (\{m_1\},\{m_2\}) \\
    addr  =  15: & (\{m_1\}, \{m_3\})
\end{array}
\]
In this example, the potential correlation between $m_2$ and $m_3$ in the original trace is removed in the sliced traces, thus avoiding the false pattern between those two messages to be mined.

\begin{algorithm}[tb]
\caption{\textbf{FlowMiner}}
\label{algo:flowminer}
\textbf{Input:} A set of traces $T$\\
\textbf{Output:} {Set of patterns $M$}\\
\SetAlgoNoLine
$T' := TraceProcessing(T)$\;
$(C, R) := Mining(T')$\;
$(C, R) := Chaining(C, R)$\;
$PostProcessing(C)$\;
$PostProcessing(R)$\;
$M = C\cup R$\;
\end{algorithm}

\begin{algorithm}[tb]
\caption{{\bf Mining}}
\label{algo:mining}
\textbf{Input:} A set of traces $T$\\
\textbf{Input:} A set of events $E$  \\
\textbf{Output:} $C, R$ with patterns of length two\\
$E := \{\mbox{messages occurred in } T'\}$\;
$C := \emptyset$\;
$R := \emptyset$\;
\ForEach{$m_1, m_2 \in E$}{
        \If{$causal(m_1, m_2)\ \wedge\ \mathit{conf_f}(m_1, m_2, T) = 1$}{
            $C := C \cup (m_1, m_2)$\;
        }
        \If{$causal(m_1, m_2)\ \wedge \mathit{conf_b}(m_1, m_2, T) = 1$}{
             $R := R \cup (m_1, m_2)$\;
        }
    }
\end{algorithm}

\subsection{Mining Patterns}
\label{sec:mining}

Algorithm \ref{algo:mining} implements function $\mathit{Mining}$ in Algorithm~\ref{algo:flowminer}.  It takes a set of traces $T$, and returns two sets of binary patterns such that patterns in $C$ and $R$ have 100\% forward and backward confidence, respectively.  
These binary patterns are building blocks to construct longer patterns. During the mining process, two messages $m_1$ and $m_2$ that satisfy the condition $\mathit{causal}(m_1, m_2)$ are considered as candidate for patterns. Otherwise, they are ignored even though the confidence between them is $100\%$.  Applying the causality checking leads to great reduction in the number of invalid patterns that could be mined otherwise. 
See Definition~\ref{causal} for definition of the structural causality.\\


Algorithm~\ref{algo:mining} takes a set of traces as input. The more diverse the traces are in terms of how flows instances are interleaved, the better the mining results are in terms of precision.  More specifically, two messages that have different temporal orderings in different traces are not considered as patterns. This is because we aim to mine patterns are that invariant over different execution runs.    

The forward and backward confidence calculations in Definitions~\ref{fconf} and \ref{bconf} are extended to consider a set of traces as follows.  Let $m_1$ and $m_2$ be two messages.  For each trace $\rho$, if  $\mathit{supp}(m_1) = 0$, the forward confidence of $m_2$ given $m_1$ on $\rho$, $\mathit{conf_f}(m_1, m_2, \rho)$, is undefined.  Then, the forward confidence of $m_2$ given $m_1$ over a set of traces, $\mathit{conf_f}(m_1, m_2, T)$ is $\mathit{conf_f}(m_1, m_2, \rho)$ averaged over all traces $\rho \in T$ where it is defined. The backward confidence over a set of traces can be calculated similarly.





\begin{algorithm}[tb]
\caption{\textbf{Chaining}}
\label{algo:chaining}
\textbf{Input:} Candidate pattern sets $C$, $R$\\
\textbf{Output:} {Longer pattern sets $C$, $R$ chained using three chaining cases}\\

\SetAlgoNoLine
\While{$|C|$ increases}{
\ForEach{$p_1, p_2 \in C$}{
    \If{$p_1 = A\#S \mbox{ and }  p_2=P\#B \mbox{ and }  S=P$}{ 
        $C := C \cup p_1 \# B$ 
}
}
}


\While{$|R|$ increases}{
\ForEach{$p_1, p_2 \in R$}{
    \If{$p_1 = A\#S \mbox{ and } p_2=P\#B \mbox{ and } S=P$}{ 
        $R := R \cup p_1 \# B$ 
}
}
}
$ R' := \emptyset$\;
\ForEach{$p_1 \in R, p_2 \in C$}{
        \If{$p_1 = A\#S \mbox{ and } p_2=P\#B \mbox{ and } S=P$}{ 
            $R' = R' \cup p_1 \# B$

            }
            }
$R := R'$\;

\ForEach{$p_1 \in C_{Trans}, p_2 \in R_{Trans}$}{
    \If{$p_1 = A\#S \mbox{ and } p_2=P\#B \mbox{ and } S=P$}{ 
        \If{$\mathit{conf}_f(\mathit{first}(p_1), last(p_2)) =~100\%$} 
        {
            $C := C \cup p_1 \# B$ 
        }
        \If{$\mathit{conf}_b(\mathit{first}(p_1), last(p_2)) =~100\%$} 
        {
            $R := R \cup p_1 \# B$ 
        }
    }
}
\end{algorithm}

\subsection{Chaining Patterns}

Although binary patterns are interesting, they only provide limited information on SoC executions. In order to gain more comprehensive understanding about the SoC executions, which is beneficial for debugging, it is necessary to extract longer patterns.
A chaining rule is described in \emph{Perracotta} \cite{Yang06perracotta:mining} that constructs longer patterns by combining binary patterns.  Specifically, for patterns $(A, B)^\ast$, $(B, C)^\ast$, $(A, C)^\ast$, they can be combined as $(A, B, C)^\ast$. However, this approach is only applicable to patterns of events that are strictly alternating.  This situation rarely exists in SoC traces where messages from different flows are typically interleaved.  Therefore, its effectiveness is limited.  Consider the following example for an illustration.
\[
\begin{array}{ll}
\rho_0: & \{a,b,c,a,b,c\} \\
\rho_1: & \{a,a,b,c,b,c\}     
\end{array}
\]
The above two traces can be viewed as being generated by executing the pattern $(a, b, c)$ twice in each run. In the first run, the instances of that pattern are executed one after the other.  In the second run, both instances are executed concurrently.  The chaining rule in \emph{Perracotta} is able to discover pattern $(a,b,c)^\ast$ from trace $\rho_0$ as such pattern alternates in $\rho_0$.  However, it fails to find such a pattern from $\rho_1$ as it is unable to mine $(a,b)$, $(a,b)$, $(a,b)$ for chaining since those binary patterns are not alternating.

In \emph{FlowMiner}, several rules, as shown in Algorithm~\ref{algo:chaining}, are developed for chaining patterns to form longer ones.  They are explained as follows. 

\begin{figure}[tb]
\centering
\begin{tabular}{ccc}
\includegraphics[height=1.5in]{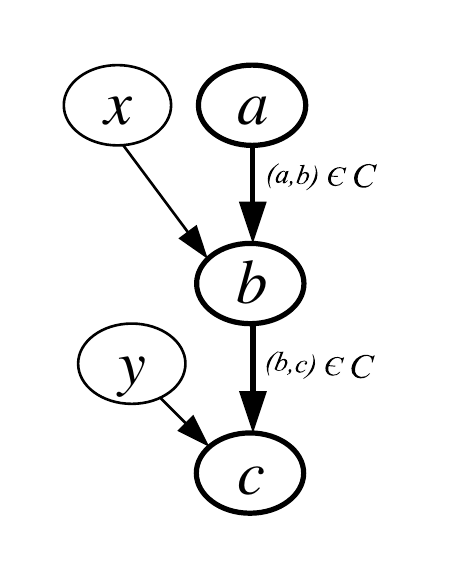} & \hspace{-20pt} \includegraphics[height=1.5in]{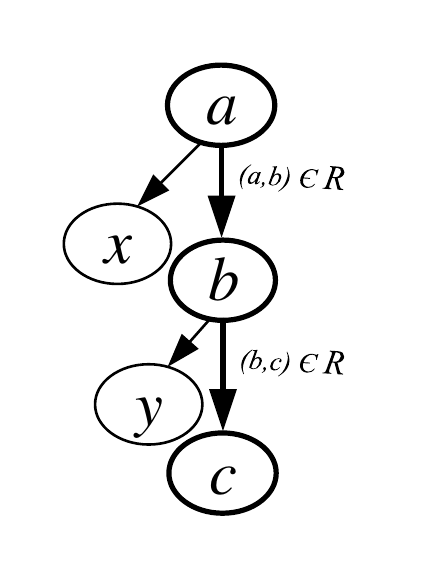} & \hspace{-20pt} \includegraphics[height=1.5in]{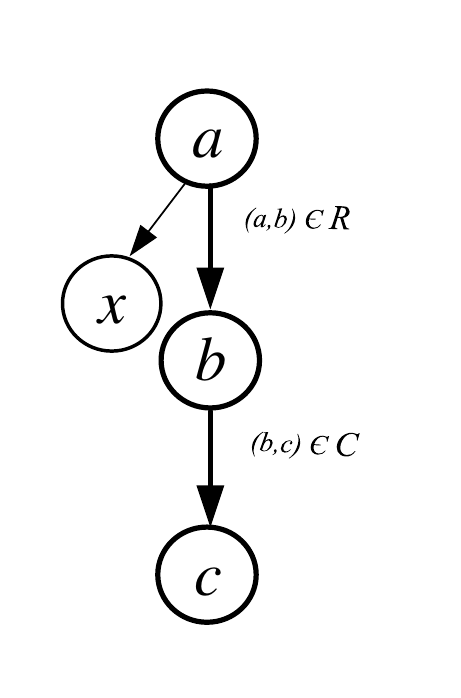} 
\\
(a) & (b) & (c)
\end{tabular}
\caption{Different flow examples to illustrate differnt chaining rules.}
\label{fig:chaining}
\end{figure}


\begin{enumerate}
    \item The first rule (line 3--6 in Algorithm~\ref{algo:chaining}) chains two patterns from set $C$ to form longer patterns.  For two patterns $p_1$ and $p_2$ to be chained, they need to share some common sub-sequence as prefix and suffix, respectively. For example in Fig.~\ref{fig:chaining} Flow 1, from the trace resulting from executing the pattern, binary patterns $(a, b)$, $(b, c)$ can be mined with $100\%$ forward confidence. They can be chained to obtain a longer pattern $(a,b,c)$.  Suppose that the pattern resulting from chaining rule~1 is $p = (m_0, \ldots, m_n)$. This pattern has the following property.
    \[
        \forall{0 \leq i \leq n-1},\ conf_f(p[0,i], p[i+1, n])= 100\%.
    \]
    Therefore, this pattern is added into set $C$ for further chaining.  This chaining rule is repeatedly applied until no patterns can be chained and added to $C$. 
  
  \item The chaining rule 2 (line 7--10 in Algorithm~\ref{algo:chaining}) is similar to the rule~1 except that it considers patterns from the set $R$.  As illustrated in Fig.~\ref{fig:chaining} Flow 2, from the trace resulting from executing the pattern, binary patterns $(a,b)$ and $(b, c)$ can be mined with the backward confidence of $100\%$.  Combining these two patterns leads to a new pattern $(a, b, c)$.  Suppose that the pattern resulting from chaining rule~2 is $p = (m_0, \ldots, m_n)$.  Similarly, the following property holds for the pattern $p$ as the result of this rule.
  \[
        \forall{0 \leq i \leq n-1},\ conf_b(p[0,i], p[i+1, n])= 100\%.
    \]
    The pattern $p$ is added to the set $R$.  This chaining rule is repeatedly applied until no patterns can be chained and added to $R$. 
   
  \item The chaining rule 3 (line 11--15 in Algorithm~\ref{algo:chaining}) explores additional opportunities of chaining pattern from patterns in the set $R$ to those in the set $C$. 
  This rule is illustrated in Fig.~\ref{fig:chaining} Flow 3.  From the trace resulting from executing that pattern, patterns $(a,b)$ and $(b, c)$ can be mined with the backward and forward confidence of $100\%$, respectively. After chaining, the resulting pattern $(a, b, c)$ satisfies $conf_b(a, (b,c)) = 100\%$.
   
  Suppose that $p = (m_0, \ldots, m_n)$ is the result of chaining $p_1$ and $p_2$ where $p_1 \in R$ and $p_2 \in C$ using the rule~3.  Also suppose that $p_2 = (m_k, \dots, m_n)$. Then, the following property holds for $p$.
  \[
        \forall{0 \leq i \leq k-1},\ conf_b(p[0,i], p[i+1, n])= 100\%.
    \]
    Therefore, the resulting pattern is added into set $R$.
\end{enumerate}


\subsection{Evidence-Oriented Chaining}  
\begin{figure}[tb]
  \vspace{-20pt}
  \begin{center}
    \includegraphics[width=0.2\textwidth]{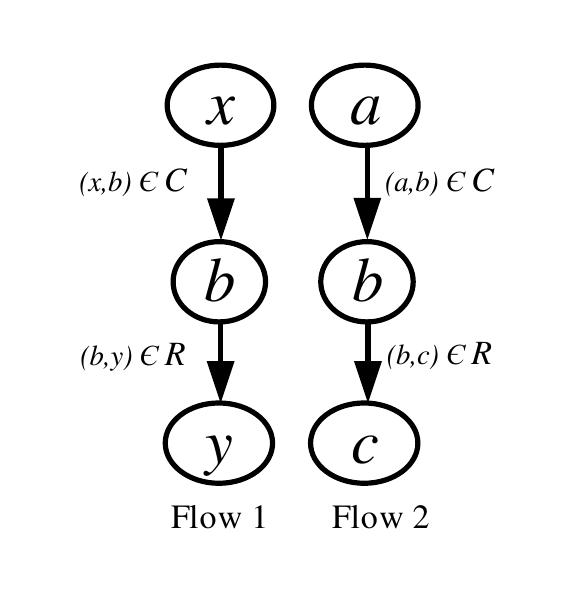}
  \end{center}
  \vspace{-20pt}
  \caption{An example of two flows that share a common message}
  \label{fig:chain4}
  \vspace{-8pt}
\end{figure}

In the previous section, we do not consider chaining patterns in set $C$ to patterns in set $R$ as we are not able to show the property of the resulting patterns similar to those for the chaining rules described above.  However, there are situations that require such chaining.  Consider the flows in Fig.~\ref{fig:chain4} as an example.  These two flows are simple, however, they share a common message $b$.  From the trace resulting from executing these two flows, the following patterns can be mined 
$C = \{(a,b), (x,b)\}$ and $R=\{(b,c), (b,y)\}$.  The chaining rules described in the last section cannot chain them together, thus missing the longer patterns in the original specification.   

To address that limitation, we develop a new chaining rule (line 17--20 in Algorithm~\ref{algo:chaining}) to obtain more interesting and longer patterns combining patterns from $C$ to those in $R$.  The idea of this new rule is that whether a pattern in $C$ can be chained with another pattern in $R$ depends on whether the chaining result can explain some previously mined pattern. 
Specifically, given a pattern $p_1 \in C$, and $p_2\in R$ such that they share some common sequence as suffix and prefix, respectively, if $conf_f(first(p_1), last(p_2)) = 100\%$, then they are chained together to form a longer pattern. The reasoning behind this rule is as follows.  If we know that $first(p_1)$ leads to $last(p_2)$ in the future for certain, then the pattern by chaining $p_1$ and $p_2$ is an evidence to support such observation.  For the example in Fig.~\ref{fig:chain4}, this rule can successfully mine the two original flows.  The similar case exists when $conf_b(first(p_1), last(p_2)) = 100\%$.



Suppose that pattern $p=(m_0, \ldots, m_n)$ is the result from using this evidence oriented chaining rule. 
Then, the following property holds for $p$.
\[
    \mbox{if } \mathit{conf_f}(p[0], p[n]) = 100\%, then~ \mathit{conf_f}(p[0,n-1], p[n])= 100\%,
\]
and
\[
    \mbox{if } \mathit{conf_b}(p[0], p[n]) = 100\%, then~\mathit{conf_b}(p[0], p[1,n])= 100\%,
\]

When the first three chaining rules are used, the number of patterns may be reduced as the shorter patterns become redundant with respect to the chained patterns, and thus are removed.  However, the new chaining rule always increases the total number of mined patterns.  One undesirable impacts would be the large number of invalid patterns mined as the result of one invalid binary pattern.  In this situation, user inputs would be invaluable to identify invalid binary patterns, and thus avoiding the evidence patterns to be generated from the beginning.  In general, users are more capable of determining the validity of binary patterns than the longer ones.      

\section{Experimental Results}

In this section, we evaluate \emph{FlowMiner} on two different types of traces: synthetic traces and simulation traces of the RTL model of an SoC prototype.  The purpose of synthetic traces is to provide a well specified evaluation environment that allows us to gain a better understanding of the strengths and limitations. The RTL simulation provides a more realistic environment for evaluation.  

In order to evaluate the proposed mining framework, we compare \flowminer with \perracotta~\cite{Yang06perracotta:mining} which is a well-known software API mining tool.
\perracotta mines binary alternating patterns from  traces that are sequences of software API calls.  It also introduces a chaining rule that allows longer alternating patterns to be derived from the shorter ones. We use the code hosted on GitHub~\cite{noauthor_modelinference}. However, that code does not include implementation of the chaining rule.  
We re-implemented the chaining rule as described in~\cite{Yang06perracotta:mining}.  \perracotta returns patterns with different satisfaction rates, the concept of which is similar to that of confidences defined in this paper.  For the comparison, this paper only considers the patterns returned by \perracotta with $100\%$ satisfaction rate.  Readers are referred to \cite{Liu:2013} for more details about \perracotta.  

We use {\bf precision} and {\bf recall} as the performance metrics.  {Precision} is defined as the ratio between the number of valid patterns mined versus the total number of mined patterns.
{Recall} is defined as the ratio between the number of mined \emph{ground truth patterns}\footnote{These are mined patterns that exactly match some ground truth patterns} versus the number of ground truth patterns available. We try to get high recall with a corresponding high precision because if the precision is not high, mined patterns are largely invalid.

\subsection{Experiments on Synthetic Traces}

In the first set of experiments, we evaluate the \emph{FlowMiner} over synthetic traces, which provide a clearly defined with known end results for more accurate assessments.  These synthetic traces are generated based on ten flows which consist of $64$ sequential patterns.  These flows are adapted and simplified from cases implemented in industry SoC designs.  They include cache coherence protocols, CPU downstream read/write, peripheral upstream read/write, power management, etc.  The sequential patterns used for synthetic trace generation are denoted as \emph{ground truth} (GT).

To generate synthetic traces, $10$ instances of each of $64$ patterns are created and put in a pool.  Then, the patterns in the pool are executed in three different ways, resulting in three different sets of synthetic traces.  For the first trace set, a pattern is randomly selected from the pool, and executed atomically to the end before another pattern is selected for execution.  In the resulting traces, there is no interleaving among different patterns, and there is a single message in each step of a trace.  This trace set is referred to as \emph{single-event, non-interleaving}.  The second trace set, multiple patterns can be active at the same time, however, we still restrict a single message in each step.  It is referred to as \emph{single-event, interleaved}.  The third trace set is generated by executing multiple patterns simultaneously, and multiple messages may occur in a single step.  This set is referred to as \emph{multi-event, interleaved}.  Each trace set includes $100$ traces generated with random orderings among the patterns.   

We apply \flowminer and \perracotta to those three trace sets, and results are reported in Table~\ref{tab:interleaved}.  The objective is to evaluate the above two approaches by comparing the mined patterns against those $64$ ground truth patterns in terms of precision and recall.

\begin{figure}[tb]
\begin{center}
\includegraphics[ height=1.75in,width=3in,angle=0]{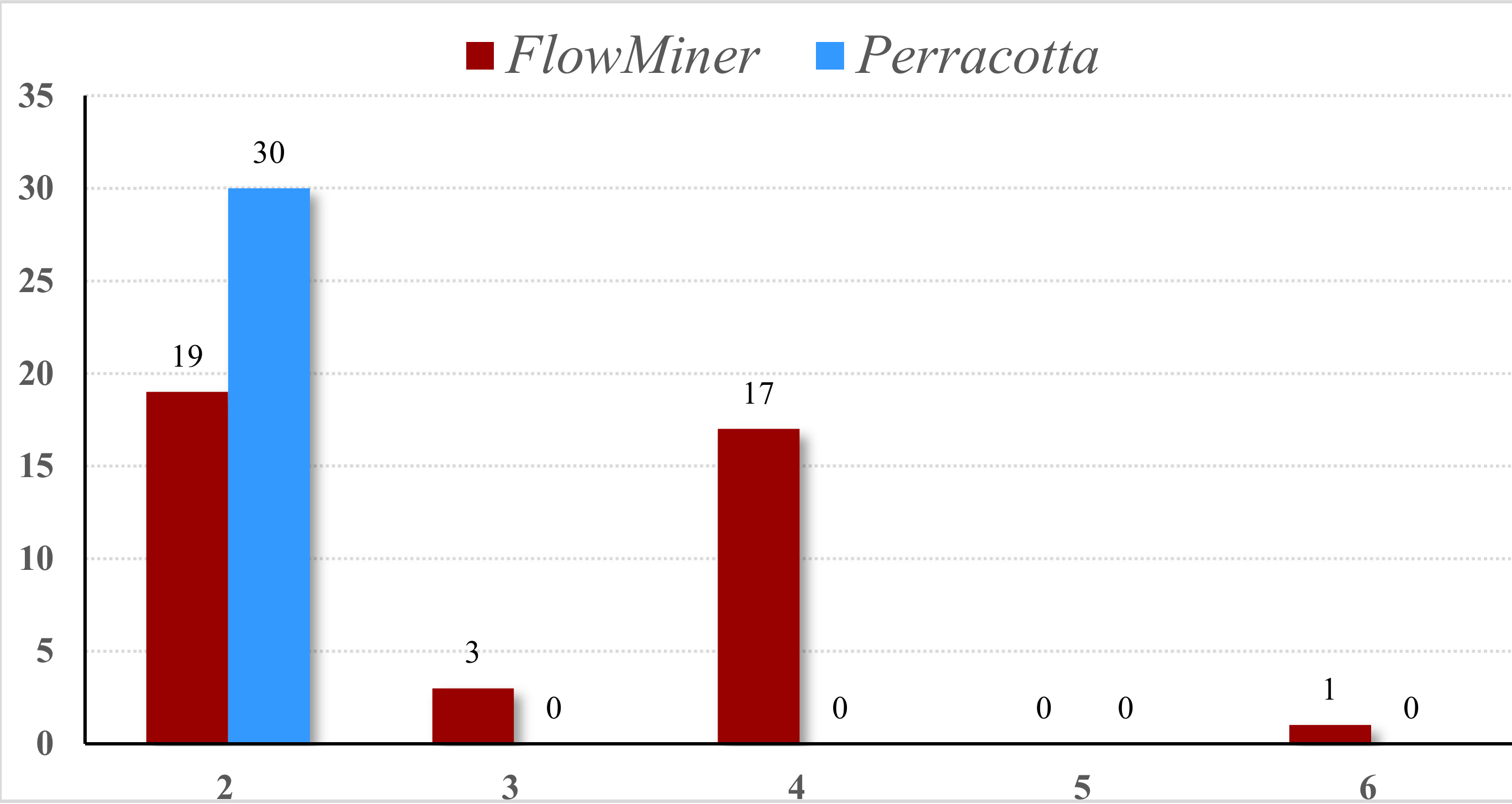}
\caption{The length distributions of mined patterns with \flowminer and \perracotta from single-message non-interleaved traces.}
\label{fig:patt_distribution_non_inter}
\end{center}
\end{figure}

\medskip
\noindent \textbf{Single-message, non-interleaved traces}. The experimental results for this set of traces are  shown in row~1 of Table~\ref{tab:synthetic-results}.  Both approaches perform well in terms precision since the messages in the traces are properly ordered, thus leading to great reduction in the false correlations among messages.  However, their performance is not as good in  recall scores, which indicates that many ground truth patterns are not mined with either approach.  In this respect, \flowminer still does better than \perracotta by mining $10$ more valid patterns.   Fig.~\ref{fig:patt_distribution_non_inter} shows the distributions of mined patterns in terms of their lengths.  All the patterns mined with $Perracotta$ is binary, while \flowminer mines many more longer patterns, which are more interesting in SoC validation.  
Note that $FlowMiner$ removes short patterns after they are used to construct longer patterns with Algorithm~\ref{algo:chaining}.
Therefore, the number of binary patterns found with $FlowMiner$ is actaully larger than what is shown in Fig.~\ref{fig:patt_distribution_non_inter}.

\medskip
\noindent\textbf{Single-message, interleaved traces.} 
The results for this set of traces are  shown in row~2 of Table~\ref{tab:synthetic-results}.  The comparison is striking as \flowminer can mine over $100$ patterns while \perracotta mines only two.  As indicated above, unlike software traces that \perracotta targets,  SoC traces are highly concurrent with many patterns interleaved or occurred simultaneously during execution. As a result, \perracotta is not able to extract many valid patterns from SoC traces effectively as the alternating patterns impose a stronger condition on qualification of sequences as patterns.

\begin{figure}[tb]
\begin{center}
\includegraphics[height=1.75in,width=3in,angle=0]{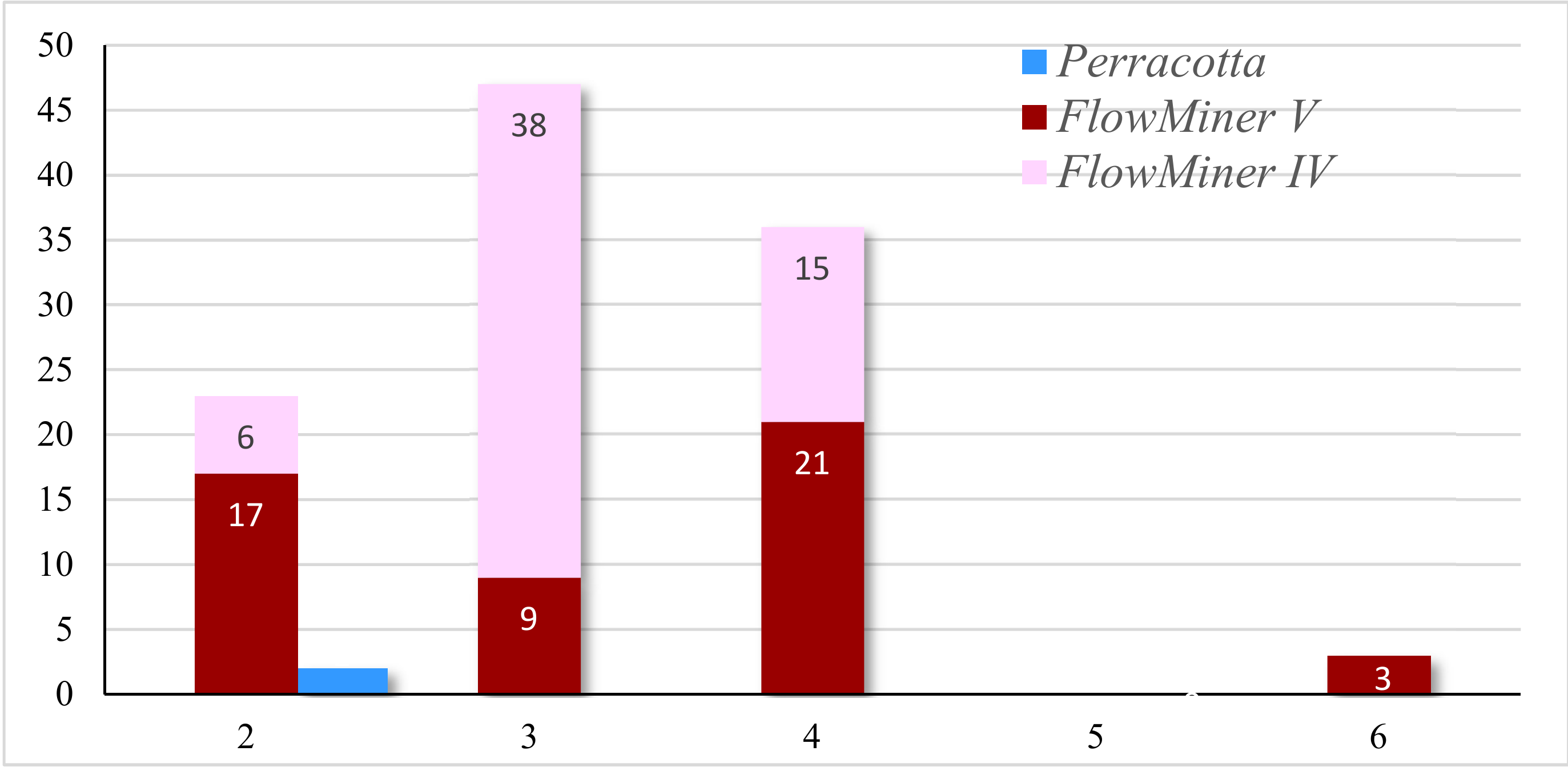}
\caption{The length distributions of mined patterns with \flowminer and \perracotta from single-message interleaved traces.  For the bars corresponding to mining results with \flowminer, the dark red segments show the numbers of valid patterns, while the light pink segments show the numbers of invalid patterns. In the figure, \emph{V} and \emph{IV} mean valid and invalid, respectively.}
\label{fig:patt_distribution_inter}
\end{center}
\end{figure}

The precision with $FlowMiner$ is worse.  This indicates that a large number of invalid patterns mined from the interleaved traces. Interleaving messages during a SoC execution can introduce temporal orderings among messages that do not exist in the non-interleaved case, and these temporal orderings cause \flowminer to mine those invalid patterns.  On the other hand, the strong condition of the alternating patterns mined with \perracotta can avoid many of those invalid patterns, in this case all of them.  In terms of recall, \flowminer performs better by finding the same number of the ground truth patterns, while \perracotta finds none.  
Fig.~\ref{fig:patt_distribution_inter} shows the distributions of mined patterns by both of the tools in terms of different lengths.  This figure shows the capability of \flowminer in finding longer patterns from the interleaved traces.  

It is interesting to notice that the $FlowMiner$ finds more valid patterns from the interleaved traces than from non-interleaved traces.  An explanation is that the message interleavings may introduce additional temporal orderings compared to the non-interleaved case.  These additional temporal orderings may cause certain patterns belonging to branching structures of flows to be mined.  A side effect of these additional temporal orderings is the larger number of invalid patterns to be found as well. 


\begin{table}[!t]
\renewcommand{\arraystretch}{1.3}
\caption{Mining from synthetic traces}
\label{tab:synthetic-results}
\centering
\begin{tabular}{ p{2cm}| c c c c c }
\hline 
\bfseries Traces & \bfseries Tool & \bfseries \vtop{\hbox{\strut \#Pattern }\hbox{\strut mined}} & \bfseries Precision & \bfseries Recall \\

\hline 
\hline
\multirow{2}{*}{\vtop{\hbox{\strut single-message, }\hbox{\strut non-interleaved}}} & $FlowMiner$ & 40 & 100\% & 18.75\% \\
 & $Perracotta$ & 30 & 100\% & 6.25\% \\
 

\hline
 
\multirow{2}{*}{\vtop{\hbox{\strut single-message, }\hbox{\strut interleaved}}} & $FlowMiner$ & 109 & 46\% & 18.75\%\\
 & $Perracotta$ & 2 & 100\% & 0.00\% \\
 \hline
 
\multirow{2}{*}{\vtop{\hbox{\strut multi-message, }\hbox{\strut interleaved}}} & $FlowMiner$ & 114 & 43\% & 18.75\%\\
& & & & \\
\hline
\end{tabular}
\end{table}




\begin{table}[!t]
\renewcommand{\arraystretch}{1.3}
\caption{Mined patterns of different length from multi-message synthetic traces}
\label{tab:multi_event_synthetic}
\centering
\begin{tabular}{c c c c c c c |c }
\hline
\bfseries Length & \bfseries 2 & \bfseries 3 & \bfseries 4 & \bfseries 5 & \bfseries 6  & \bfseries Total & \bfseries GT\\
\hline \hline
V\&F & 17 & 14 & 17 & 0 & 1  & 49 & 12\\
IV\&F& 6 & 52 & 7 & 0  & 0 & 65 & \\
\hline
\end{tabular}
\end{table}

\medskip
\noindent \textbf{Multi-message, interleaved traces.} This set of traces intends to simulate real SoC execution scenarios. 
In these traces, there may be multiple messages in each step of execution, and multiple patterns can be active at a time.  
\perracotta is not able to handle these traces, thus only results with \flowminer are shown in row~4, Table~\ref{tab:synthetic-results}.  Table~\ref{tab:multi_event_synthetic} shows the breakdown of the results in terms of length and validity of mined patterns.  These results show that concurrency and interleaving together make mining valid and longer patterns a very challenging task where a lot of invalid patterns are often mined.  For this reason, the corresponding precision is worse than the precision values in the other two cases.  On the other hand, the same number of ground truth patterns, which is $12$, is still found. 

\subsection{Experiment on Simulation Traces}

\begin{figure}[tb]
\begin{center}
\includegraphics[height=1.75in,width=3in,angle=0]{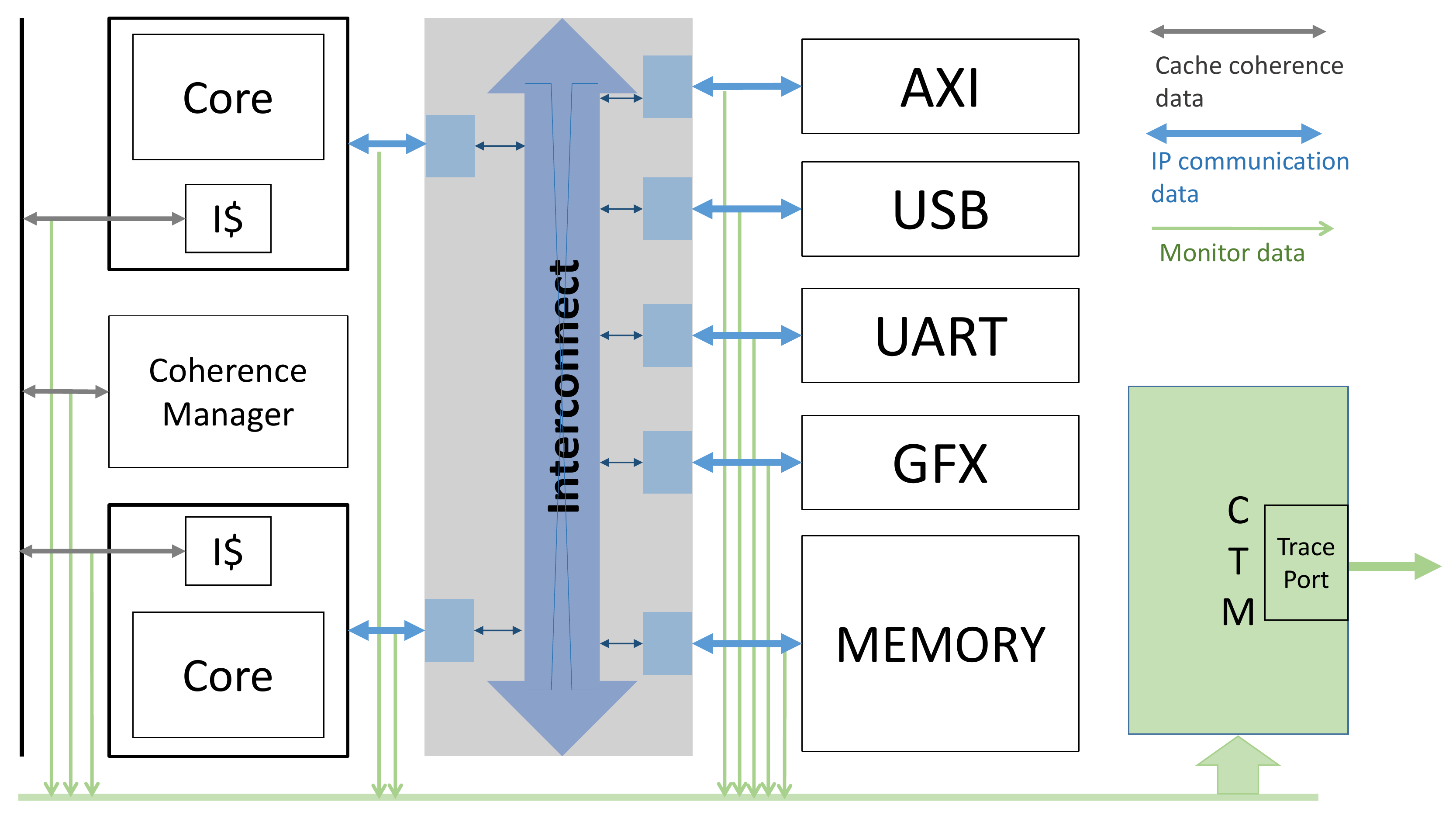}
\caption{An SoC prototype where each communication link is attached with a monitor. There can be multiple links between a pair of components.}
\label{soc}
\end{center}
\end{figure}

In this experiment, we evaluate \emph{FlowMiner} using the traces generated by simulating a non-trivial SoC design model \cite{cao_mining_2020}, as shown in Fig.~\ref{soc}.  This model has a typical structure of a realistic SoC designs.  It contains two CPU cores with private caches. They are connected with a cache coherent bus.  There are a number of peripheral IP blocks.  All the CPU cores and peripheral blocks are connected a switch-based on-chip network, which can concurrently execute multiple flows simultaneously.  This model is specified in RTL VHDL. 

The on-chip network of this SoC model implements the $10$ flows used for synthetic trace generation.  A random test environment is created to simulate the SoC model where CPUs and the peripheral blocks are configured to randomly select a flow to initiate with a random delay between 1 to 10 cycles.  The traces from the RTL simulation are obtained, and then abstracted to traces of messages. The difference between these simulation traces and those synthetic ones is that the traces from simulation include runtime information, specifically memory addresses.  We would like to find out whether such runtime information can help to improve the mining performance of \emph{FlowMiner}.  Another difference is that the ground truth patterns embedded in the individual simulation traces are unknown as the patterns are randomly selected during simulation.  In order to calculate the precision and recall of mining results, we use the set of \emph{all} ground truth patterns implemented in the SoC design model.  However, the recall scores may be pessimistic as not all the ground truth patterns are presented in the traces. 

In each simulation run initialized with a different random seed, a total of $500$ pattern instances are randomly selected for execution.  We collect $228$ traces.  We then apply address-based slicing, and derive a set of over $130$ thousand sliced traces.  As the original traces contain multiple messages in certain steps,  Tables~\ref{tab:orig_simulation} only shows the mining results with \flowminer.  For sliced traces, both \flowminer and \perracotta are used, and mining results are shown in Table~\ref{tab:sliced_trace}.

Table ~\ref{tab:orig_simulation} shows the mined patterns with \flowminer in terms of lengths and validity.  It can be seen that there many more invalid patterns compared to the number of valid patterns.  The precision in this case is about $5\%$, which indicates the challenge of mining from high concurrent SoC traces.  On the other hand, \flowminer is able to mine more ground patterns, and the recall score ($31\%$) is better compared with synthetic experiments. 

The mining results from the sliced traces are shown in Table~\ref{tab:sliced_trace}.  It can be seend that trace slicing has a big impact on mining performance.  Compared with results in Table~\ref{tab:orig_simulation}, \flowminer is able to mine more valid patterns while avoiding all invalid patterns, therefore both precision and recall are improved. Compared with \emph{Perracotta}, \flowminer mines over twice as many patterns.  Again, this is due to the strong condition imposed by mining alternating patterns in \emph{Perracotta}. 

\begin{figure}[!htb]
\begin{center}
\includegraphics[height=1.75in,width=3in,angle=0]{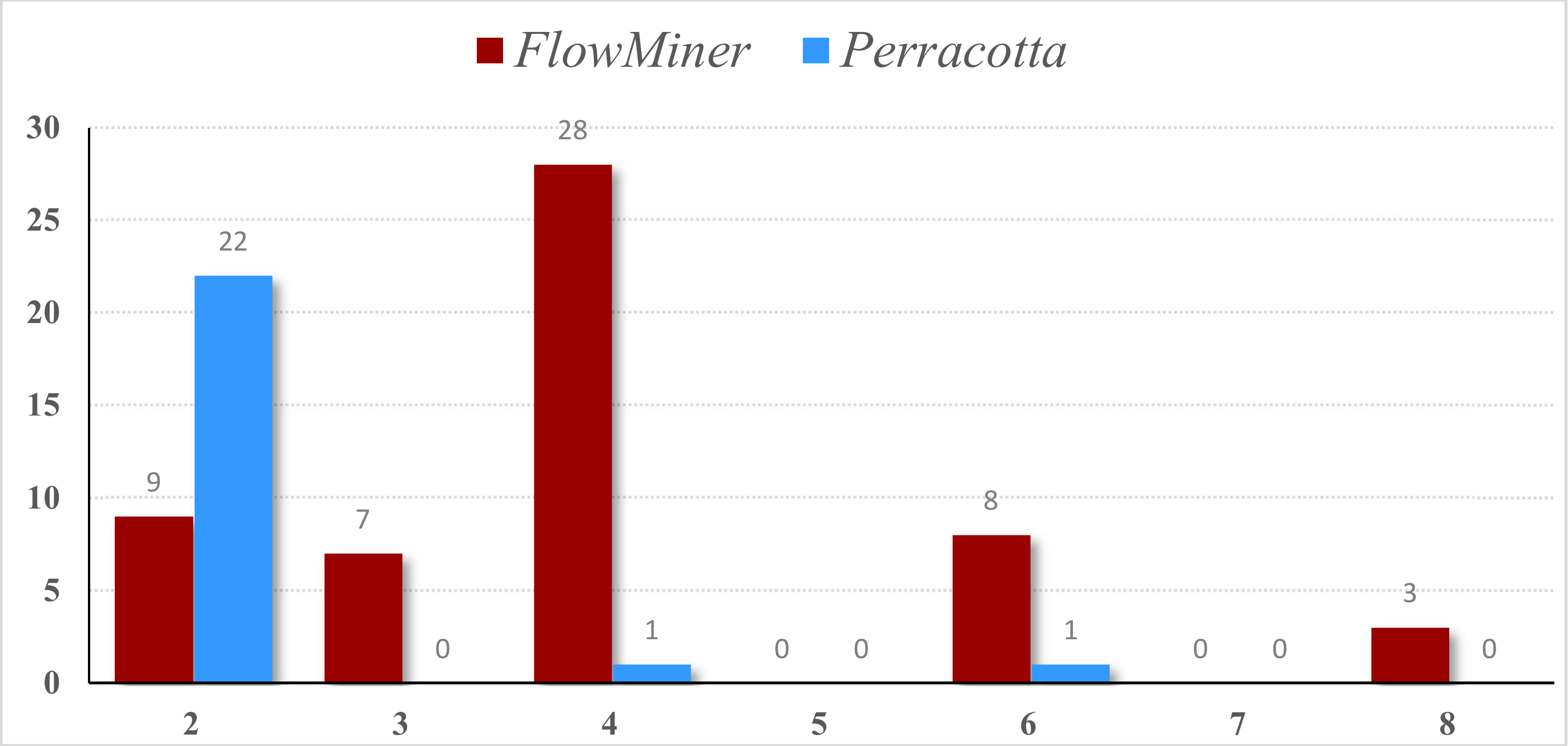}
\caption{The length distributions of mined patterns with \flowminer and \perracotta from sliced simulation traces.}
\label{fig:patt_distribution_sliced}
\end{center}
\end{figure}


\begin{table}[!t]
\renewcommand{\arraystretch}{1.3}
\caption{Mined patterns from 224(>1M messages) simulation traces.}
\label{tab:orig_simulation}
\centering
\begin{tabular}{c c c c c c c c c | c}
\hline
\bfseries Length & \bfseries 2 & \bfseries 3 & \bfseries 4 & \bfseries 5 & \bfseries 6 & \bfseries 7 & \bfseries 8 & \bfseries Total & \bfseries GT\\
\hline \hline
V\&F & 10 & 0 & 17 & 0 & 13 & 0 & 5 & 45 & 20\\
IV\&F& 4 & 12 & 26 & 79 & 213 & 259 & 265 & 859 & \\
\hline
\end{tabular}
\end{table}

\begin{table}[!t]
\renewcommand{\arraystretch}{1.3}
\caption{Mining results from the sliced traces. \flowminer mines $16$ ground truth patterns while \perracotta only mines 2.}
\label{tab:sliced_trace}
\centering
\begin{tabular}{c c c c  c}
\hline
 &  \multicolumn{1}{c}{ \bfseries \#Patterns Mined } & \multicolumn{1}{c}{ \bfseries Precision} & \multicolumn{1}{c}{ \bfseries Recall} \\
\hline \hline
$FlowMiner$ & 55 & 100\% & 25\%\\
$Perracotta$ & 24 & 100\% & 3.12\%\\
\hline
\end{tabular}
\end{table}

\subsection{Runtime Analysis}

Table 5 gives the runtime information for the experiments performed in sections~5.1 and 5.2. \perracotta is written in Java, and we use the compiled version downloaded from Github. The version we use does nbot implement the chaining rule described in~\cite{Yang:2006}, therefore we implement it in Python as described in~\cite{JinlinYang:2007}.
On the other hand, \flowminer is written entirely in Python.  All experiments are performed on a Windows PC with a quad Core-i7 processor and $16$~GB memory.  

From the table, it can be seen that \flowminer is slower than \perracotta due to the performance disadvantage of Python compared to Java.  Additionally, in \flowminer a significant amount of time is spent on post-processing where redundant patterns are removed.  If that step is removed, the runtime between \flowminer and \perracotta in all experiments are comparable.

\begin{table}[tb]
\caption{Runtime comparison between \flowminer and \perracotta in different experiments. Time is in second.  SM, MM, I, and NI refer to "single-message", "multi-message", "interleaved", and "non-interleaved", respectively.}
\label{tab:runtime}
\centering
\begin{tabular}{|c|c|c|c|}
\hline
\multicolumn{2}{|c|}{\bfseries Traces} & \bfseries Tools & \bfseries Runtime  \\ 
\hline\hline
\multirow{6}{*}{Synthetic}  & \multirow{2}{*}{SM, NI} & \flowminer & 58 \\
                            &                       & \perracotta & 14 \\
                            \cline{2-4}
                            & \multirow{2}{*}{SM, I} & \flowminer & 60 \\
                            &                       & \perracotta & 13 \\
                            \cline{2-4}
                            & \multirow{2}{*}{MM, I} & \flowminer & 31 \\
                            &                       & \perracotta & $-$ \\ 
                            \hline\hline
\multirow{4}{*}{Simulation} & \multirow{2}{*}{Original} & \flowminer & 210 \\
                            &                       & \perracotta & 20 \\
                            \cline{2-4}
                            & \multirow{2}{*}{Sliced} & \flowminer & 124 \\
                            &                       & \perracotta & 20 \\
                            \hline
\end{tabular}
\end{table}



 
 


\section{Discussions}

In this section, we discuss a few observations about \emph{FlowMiner}.

\paragraph{\bf Mining Complex Patterns.} \emph{FlowMiner} can discover strong temporal relations among messages observed in a set of traces. It uses $100\%$ forward and backward confidence to mine assertions which hold over all traces as we are interested in mining message flows implemented in SoC communication fabrics. The analysis of experimental results indicates that these two confidence measures, used together, allow flows with branching structures to be fully extracted from traces if those flows do not share common messages in their specifications. 

On the other hand, if multiple flows with branching structures and common messages are executed, then branches of those flows may not be completely mined.  One examples includes two flows as shown in Figure~\ref{fig:ex} such that one is memory write for {\tt CPU\_0} ({\tt x=0}), while the other is memory write for {\tt CPU\_1} ({\tt x=1}).  These two flows share messages $m_5$ and $m_6$ for memory access.  The pattern $(m_4, m_5, m_6, m_7)$ may not be mined by \emph{FlowMiner} even though $(m_5,m_6)$ can be extracted.  A typical technique for mining more patterns including those branching ones is to lower the confidence level.  Its drawback is the rapid increase in the total number of mined patterns, majority of which are invalid. Additionally, finding a confidence level that balances the numbers of valid and invalid patterns can be tricky.  For mining assertions, we focus on developing the pattern chaining techniques instead of manipulating confidence levels in this work.

\paragraph{\bf Invalid Patterns.} Mining excessively large number of invalid patterns is a serious issue to limit specification mining in practical use.  The main reason for mining invalid patterns is the incorrect correlations among messages in traces. \emph{FlowMiner} uses address-based trace slicing and structural causality to exclude incorrect correlations from mining in a certain degree.  More such techniques would be beneficial to improving mining precision.    



\paragraph{\bf Traces for Mining.} \emph{FlowMiner} can take the any number of traces of different lengths provided as input for mining.  More patterns would be mined from longer traces as those traces may include more information about different flow executions.  The number of input traces can have an impact on mining precision as indicated in Section~\ref{sec:mining}.  Since temporal relations among messages extracted from traces are used approximate the causality relations, more input traces with diverse temporal orderings of flow executions can help to eliminate temporal relations that do not hold for all traces, thus reducing invalid patterns to be mined.  

This work considers lossless traces, \ie~all messages of executed flows are captured in traces.  To deal with traces with missing messages,  we can lower the confidence levels.  The side effects of that approach are explained above.  Since \emph{FlowMiner} mines assertions, the patterns that involves missing messages cannot be mined.  In simulation environment, we think this may not be an issue.  A message can be missing due to 1) a design bug, or 2) incorrect instrumentation.  In the former case, a message may be missing only in a few spots in a trace.  On the other hand, a message will be always missing if the instrumentation for observation is set up incorrectly.  When the mining results are returned to the user for examination, the user can query the database of the mined patterns to filter out patterns that are deemed invalid, and to search for patterns that are expected to hold.  If an expected pattern is not included, it signals a problem in the simulation environment.  According to the above observation, the confidence levels of the sequence for a missing pattern can reveal the potential cause. If confidence levels of a missing pattern\footnote{A missing pattern refers to a sequence that is supposed to be pattern but is not mined due to missing messages.} are higher than 0, it is likely caused by a design bug.  Otherwise, incorrect instrumentation would be a more likely problem. 
 


\section{Conclusion}

Developing an effective specification miner is a challenging endeavor.  \emph{FlowMiner} is such an attempt that targets mining patterns about communication behavior over communication fabrics of SoC designs.  It takes SoC execution traces that are highly concurrent, and produces a set of patterns, collectively to characterize a message flow specification that describes how various components of an SoC design coordinate with each other to realize system level functions. 

We are encouraged by the initial experimental results compared with two similar methods.  For future improvements in terms of higher precision and recall, more effective chaining techniques and mining methods are needed. In addition, we plan to experiment \emph{FlowMiner} on more diverse and realistic SoC designs and find out what and how domain specific information can help to achieve better mining results. 


\section*{Acknowledgment}

The research presented in this paper was partially supported by gifts from the Intel Corporation, and a grant from Cyber Florida.







\end{document}